\documentstyle[preprint,aps]{revtex}

\begin{document}
%\draft
\preprint{KAIST-TH 00/3}
\title{
Effects of flavor conserving CP violating phases in SUSY models 
}

\author{P. Ko}

\address{
Department of Physics, KAIST, 
Taejon 305-701, KOREA \\E-mail: $^a$ pko@muon.kaist.ac.kr}

%\author{Elsie Tan,  Jessie Tan and R. Sankaran}

%\address{World Scienfitic Publishing Co Ltd, 
%57 Shelton Street, Covent Garden, London WC2H 9HE, England\\
%E-mail: wspc@wspc.ox.uk}  


\maketitle

%\abstracts{
\begin{abstract}
I summarize our recent works on the effects of flavor conserving CP violating 
phases  in SUSY models on $B$ and $K$ phenomenology.
\end{abstract}
%}

\narrowtext
\tighten
\section{Introduction}
%\subsection{Motivations}\label{subsec:prod}
The minimal supersymmetric standard model (MSSM) has many 
CP violating (CPV) phases beyond the KM phase in the standard model (SM). 
These SUSY CPV phases, depending on their flavor structures, are strongly 
constrained by $\epsilon_K$ or electron/neutron electric dipole moment (EDM),
and have been considered very small ($\delta \leq 10^{-2}$ for $M_{\rm SUSY} 
\sim O(100)$ GeV ).\cite{susycp} 
Another way to solve these problems is to  consider effective SUSY models,
where decouplings of the 1st/2nd generation sfermions are invoked to evade
the EDM constraints and also SUSY FCNC/CP problems.\cite{kaplan} 
In such cases, these new SUSY phases may affect $B$ and $K$ physics. 
One strong motivation for new CP violating phases beyond the KM phase 
is related with the baryon number asymmetry of the universe.  
Electroweak baryogenesis is possible in a certain region of the 
MSSM parameter space, especially for light stop 
($120~{\rm GeV} \leq m_{\tilde{t}_1} \leq 175$ GeV)
with CP violating phases in $\mu$ and $A_t$ parameters.\cite{carena}
This light stop and new CP violating phases in $\mu$ and $A_t$ parameters 
can affect $B$ and $K$ physics, although these phases are flavor conserving.
In this talk, we report our three recent works related with this  subject.
%\cite{ko1},\cite{ko2},\cite{ko3}  
\cite{ko1,ko2,ko3}
The topics covered here are the following :
the effects of $\phi_\mu$ and $\phi_{A_t}$ on $B$ physics 
in the MMSSM, and fully supersymmetric 
CP violations in the kaon system.

\section{Effects of $\mu$ and $A_t$ phases on $B$ physics
in the more minimal supersymmetric standard model (MMSSM)}

In the MMSSM we consider in this section, only the third family squarks and
charginos can be light enough to affect
$B\rightarrow X_s \gamma$ and $B^0 - \overline{B^0}$ mixing.
We also ignore possible flavor changing squark mass matrix elements
that could generate gluino-mediated flavor changing neutral current (FCNC)
process,  dicussions of which can be found in the 
literatures,\cite{cohen-B,kkl}, for example.  
Ignoring such contributions, the only source of the FCNC in our model may be
attributed to the CKM matrix, whereas there are new CPV phases coming from 
the phases of $\mu$ and $A_t$ parameters in the flavor preserving sector 
in addition to the KM phase $\delta_{KM}$ in the flavor changing sector. 

%\subsection{Chang-Keung-Pilaftsis (CKP) EDM Constraints}
Even if the 1st/2nd generation squarks are very heavy and degenerate, there
is another important edm constraints considered by Chang, Keung and Pilaftsis 
(CKP)  for large $\tan\beta$.\cite{pilaftsis}  
This constraint comes from the two loop diagrams involving stop/sbottom 
loops, and is independent of the masses of the 1st/2nd generation squarks.
Therefore, this CKP edm constraints can not be simply evaded by making the 
1st/2nd generation squarks very heavy, and it turns out that this puts a 
strong constraint on the possible new phase shift in the 
$B^0 - \overline{B^0}$ mixing.  
We scanned over the broad parameter space 
and imposed the various experimental constraints including 
$BR(B \to X_s \gamma)$.% : 
It has to be emphasized that this parameter space is larger than that in  
the constrained MSSM (CMSSM) where the universality of soft terms at the GUT 
scale is assumed.  %Especially, we will allow $m_U^2$ to be negative as well 

%\subsection{$B^0 - \overline{B^0}$ Mixing}

The $B^0 - \overline{B^0}$ mixing is generated by the box diagrams 
with $u_i-W^{\pm} (H^{\pm})$ and $\tilde{u}_i-\chi^{\pm}$ running around 
the loops in addition to the SM contribution. 
The gluino and neutralino contributions are negligible in our model.
The chargino exchange contributions to $B^0 - \overline{B^0}$ mixing is 
generically complex relative to the SM contributions, and this effect can be 
in fact significant for large $\tan\beta (\simeq 1/\cos\beta)$, since the 
chargino contribution  is proportional to $ (m_{b} / M_W \cos\beta )^2$. 
However, the CKP edm constraint puts a strong constraint for large 
$\tan\beta$ case.
The result is that the CKP edm constraint on $2 \theta_d$ is in fact very 
important  for large $\tan\beta$, and we have $| 2 \theta_d | \leq 1^{\circ}$. 
This observation is important for the CKM 
phenomenology, since time-dependent CP asymmetries in neutral $B$ decays into 
$J/\psi K_S, \pi\pi$ etc. would still measure directly three angles of the 
unitarity triangle even if $\phi_{A_t}$ and $\phi_{\mu}$ are nonzero.  
We also find that the dilepton asymmetry (proportional to 
${\rm Re} ( \epsilon_B )$) is very small as in the SM, but $\Delta m_B$ can be
enhanced as much as $60 \%$. See Ref.~\cite{ko1} for more details.

%\subsection{Direct Asymmetry in $B \rightarrow X_s \gamma$}
The radiative decay of $B$ mesons, $B\rightarrow X_s \gamma$, is
described by the effective Hamiltonian including (chromo)magnetic dipole
operators. Interference between $b\rightarrow s \gamma$ and $b\rightarrow 
s g$ (where the strong phase is generated by the charm loop via 
$b\rightarrow c\bar{c}s$ vertex) can induce direct CP violation in 
$B\rightarrow X_s \gamma$.\cite{KN}  The SM predicts a very small asymmetry
smaller than $0.5 \%$, so the larger asymmetry will be a clean signal for new 
source of CP violating phases. 
In our model, we find that $A_{\rm CP}^{b\rightarrow s \gamma}$ can 
be as large as $\simeq \pm 16\%$ if chargino is light enough, 
even if we impose the edm constraints. So this mode may be one of the good 
place for probing new CPV phases. 

%\subsection{The Branching Ratio for $B \rightarrow X_s l^+ l^-$}
Next let us next consider $R_{ll}$, the ratio of the branching ratio for 
$B\rightarrow X_s l^+ l^-$ in our model to that in the SM. 
In the presence of the new phases $\phi_{\mu}$ and 
$\phi_{A_t}$, $R_{\mu\mu}$ can be as large as 1.85, and the deviations from 
the SM prediction can be large, if $\tan\beta > 8$.   
As noticed in Ref.~\cite{kkl}, the correlation between the ${\rm Br}
( B\rightarrow X_s \gamma)$ and $R_{ll}$ is distinctly different from that 
in the minimal supergravisty case.\cite{okada} 

\section{Fully SUSY CP violation in the kaon system}

In the MSSM with many new CPV phases, there is an intriguing possibility that 
the observed CP violation in $K_L \rightarrow \pi\pi$ is fully supersymmetric 
due to the complex parameters $\mu$ and $A_t$ in the soft SUSY breaking terms 
which also break CP softly, or CP violating $\tilde{g} - q_i - \tilde{q}_j$. 
Our study on the first possibility in the MMSSM \cite{ko2}  
indicates that the supersymmetric $\epsilon_K$ 
(namely, for $\delta_{KM} = 0$) is less than $\sim 2\times 10^{-5}$,
which is too small compared to the observed value : 
$ | \epsilon_K | = (2.280 \pm 0.019) \times 10^{-3}$.
(See also Ref.~\cite{abel}.)

Although one cannot generate enough CP violations in the kaon system through
flavor preserving $\mu$ and $A_t$ phases in the MSSM, it is possible if one
considers the flavor changing SUSY CPV phases. In the mass insertion 
approximation (MIA), the folklore was that if one saturates the $\epsilon_K$
with $(\delta_{12}^d )_{LL}$, the corresponding $\epsilon^{'} / \epsilon_K$ 
is far less than the observed value. On the other hand, if one saturates 
$\epsilon^{'} / \epsilon_K$ with $(\delta_{12}^d )_{LR}$, the resulting 
$\epsilon_K$ is again too small compared to the data. Therefore one would 
need two independent parameters  
%\begin{equation}
$| (\delta_{12}^d )_{LL} | \sim O( 10^{-3} )$ and
$| (\delta_{12}^d )_{LR} | \sim O( 10^{-5} )$,
%\end{equation}
each of which has a $\sim O(1)$ phase. Recently, Masiero and Murayama argued 
that  such a large value of $ (\delta_{12}^d )_{LR}$ is not implausible in
general MSSM, e.g., if the fundamental theory is a string theory.\cite{mm} 
In their model, the large $(\delta_{12}^d )_{LR}$ is intimately related with 
the large $(\delta_{11}^d )_{LR}$, so that their prediction on the neutron 
edm is very close to the current experimental limit. 

In recent work,  we pointed out it is possible in fact to generate both
$\epsilon_K$ and ${\rm Re} (\epsilon^{'} / \epsilon_K)$ with a 
single complex number
$(\delta_{12}^d )_{LL} \sim  O ( 10^{-2} - 10^{-3} )$ with an order 
$\sim O(1)$ phase, if one goes beyond the single mass insertion 
approximation.\cite{ko3} Namely, the $\epsilon_K$ is generated by 
$(\delta_{12}^d )_{LL}$, whereas $\epsilon^{'} / \epsilon_K$ is generated 
by a flavor preserving $\tilde{s}_R - \tilde{s}_L$ transition followed by 
flavor changing $\tilde{s}_L - \tilde{d}_L$ transition. 
The former is proportinal to $m_s ( A_s - \mu \tan \beta ) / \tilde{m}^2$
where $\tilde{m}$ is the common squark mass in the MIA. {\it This induced 
$LR$ mixing is present generically in any SUSY models, if $| \mu \tan\beta|
\sim 10-20$ TeV.}
The only relevant question would be how one can have an $\sim O(1)$ phase
in the $(\delta_{12}^d )_{LL}$. For example, the gluino mass can have a CPV
phase $\phi_3$ which is flavor preserving. After we redefine the 
gluino field so that the the gluino mass parameter becomes real, the phase
$\phi_3$ will be tranferred to the $\tilde{g} - q_i - \tilde{q}_j$ vertex,
thereby generating CP violations in both flavor preserving and flavor changing
gluino mediated strong interactions.\cite{kanelast}   

If the KM phase were not zero in this scenario,
we cannot use the constraints coming from $\epsilon_K$ or $\Delta M_{B}$, 
since new physics would contribute to both $\Delta S= 2$ and $\Delta B= 2$ 
amplitudes.
In particular, even the third or fourth quadrant in the $(\rho,\eta)$ plane 
should be possible, in principle. 
More detailed discussions on these points will be
presented elsewhere.  Finally let us note that the recent 
observation on CP asymmetry in $B^0 \rightarrow J/\psi K_S$ depends on 
different CP violating parameter $( \delta_{i3}^d )_{AB}$ where $i=1$ or $2$, 
and $A,B = L$ or $R$, and is independent of the kaon sector we considered 
here in the mass insertion approximation. 

\acknowledgements
I am grateful to S. Baek, J.-H. Jang, Y.G. Kim, J.S. Lee and 
J.H. Park for enjoyable collaborations on the works presented in this 
talk, and also to W.S. Hou and H.Y. Cheng for their nice organization of 
this conference.   
This work is supported in part by grant No. 1999-2-111-002-5 from the 
interdisciplinary research program of the KOSEF and BK21 project of 
Ministry of Education.  

%\section*{Appendix}
%We can insert an appendix here and place equations so that they 
%are given numbers such as Eq.~(\ref{eq:app}).
%\be
%x = y.
%\label{eq:app}
%\ee

%\section*{References}

\end{document}